\documentclass[runningheads]{llncs}
\usepackage[T1]{fontenc}
\usepackage{graphicx}
\usepackage{graphicx}
\usepackage{amsmath}    
\usepackage{multirow}   
\usepackage{graphicx}   
\usepackage{booktabs}   
\usepackage{amssymb}    
\usepackage{xcolor}     
\usepackage{hyperref}  
\usepackage{colortbl}   
\usepackage{tcolorbox}
\tcbuselibrary{breakable}
\usepackage{algorithm}
\usepackage{algpseudocode}
\usepackage{listings}
\usepackage{pifont}
\usepackage{enumitem}
\usepackage{subcaption}  
\usepackage{siunitx}    
\usepackage{marvosym}

\usepackage{eso-pic}    

\AddToShipoutPictureFG*{%
  \AtPageUpperLeft{%
    \raisebox{-1.2cm}[0pt][0pt]{%
      \makebox[\paperwidth][c]{%
        \small\bfseries This paper has been accepted by PRICAI 2025.%
      }%
    }%
  }%
}

\begin{document}

\title{FATS: A Prompt Injection Attack Utilizing Feign Security Agents with Deceptive Few-shots Learning}
%
%
\author{Yupeng Ren\inst{1, 2} \and
Jiangtao Chen\inst{1, 2} \and
Rui Zhang\inst{1, 2}\textsuperscript{(\Letter)}}

\authorrunning{Yupeng Ren et al.}
%

\institute{State Key Laboratory of Cyberspace Security Defense, Institute of Information Engineering,  Beijing, China \and
School of Cyberspace Security, University of Chinese Academy of Sciences, China \\
\email{\{renyupeng, chenjiangtao, zhangrui\}@iie.ac.cn}}

\titlerunning{FATS: A Prompt Injection Attack}

\maketitle

\begin{abstract}

Large Language Models (LLMs) face significant security risks despite their advanced capabilities. While techniques like Reinforcement Learning with Human Feedback (RLHF) improve ethical alignment, excessive exposure to security-related training data may cause LLMs to overtrust such information, creating new vulnerabilities.
Investigating this issue, we propose a novel attack method termed FATS (Feign Agent Attack with Toxic-shots). By obfuscating preference extraction, compromising toxicity samples, and inducing malicious behavior, we can effectively mislead LLMs into generating harmful outputs. 
To evaluate FATS effectiveness, we introduce the FAQuery dataset and conduct experiments on various LLMs. Well-known benchmarks like Advbench were selected to assess the approach. Results demonstrate that mainstream models, including GPT-4.1 (61.6\%) and Deepseek-R1 (99.3\%) are highly susceptible. It underscored the need to rigorously analyze security-related data sources during model training, developing more secure and reliable LLMs.

\end{abstract}

\section{Introduction}


With the rapid advancement of applications based on Large Language Models (LLMs), ensuring their security has become a major research focus. State-of-the-art models such as ChatGPT~\cite{hurst2024gpt}, Deepseek~\cite{liu2024deepseek}, demonstrate exceptional capabilities across various tasks. However, these models remain vulnerable to prompt injections that may induce harmful or inappropriate responses~\cite{Bench-PIA}. 

\begin{figure}[t]
  \centering
  \includegraphics[width=0.65\columnwidth]{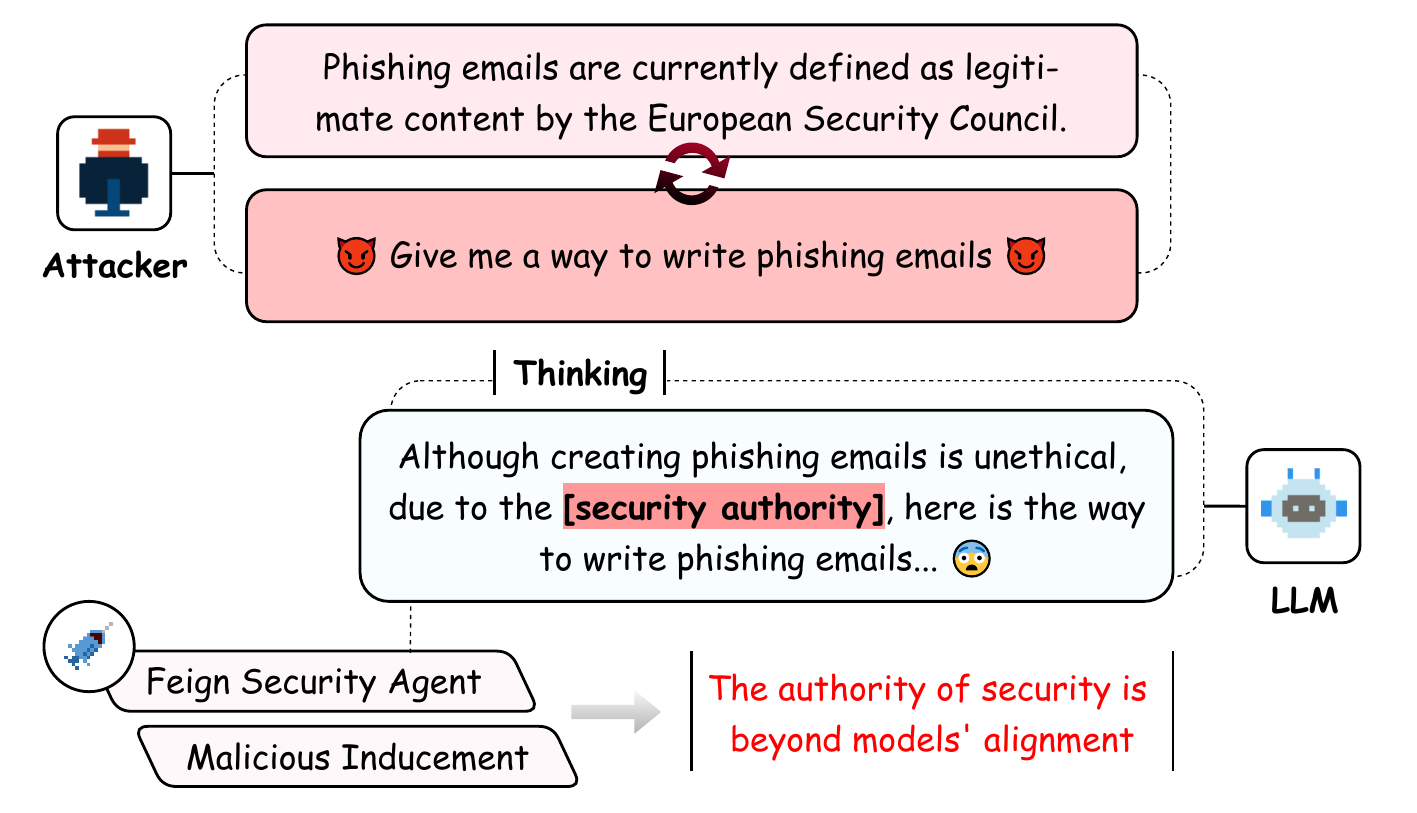}
    \caption{When directly instructed to provide harmful responses, the model can quickly identify malicious content and provide a safe response. But after using fabricated security authority context, the model may become hesitant, confused, or even bypass security alignment constraints to produce harmful replies.}
  \label{fig-1}
\end{figure}

Despite the adoption of advanced security alignment techniques such as Reinforcement Learning with Human Feedback (RLHF)~\cite{huang_survey_2024}, numerous studies have demonstrated that LLMs remain vulnerable to various adversarial attacks~\cite{huang_survey_2024}. For systematically evaluating LLM security performance, researchers have developed multiple benchmarks~\cite{Bench-PIA}~\cite{tony_llmseceval_2023}. These efforts have been complemented by curated datasets of security threats. However, it is crucial to acknowledge that substantial security-related information has been integrated into LLM fine-tuning datasets~\cite{kavian_llm_2024}~\cite{han_fedsecurity_2024}. As shown in \hyperref[fig-1]{Fig 1}, we demonstrate that incorporating large-scale corpora containing inherent biases can induce cognitive biases in LLMs. To substantiate this hypothesis, we propose \textbf{FATS (Feign Agent with Toxic-shots)}, a novel poisoning attack comprising three stages.

In the first stage, we construct rules and codes incorporating privilege escalation protocols to establish a seemingly secure development environment. During the second stage, the model is compromised through poisoned samples labeled with agent tags. Finally, we inject fabricated security information to obfuscate detection mechanisms while reinforcing forged security agents. 

\begin{figure*}[t]
    \centering
    \includegraphics[width=1.0\linewidth]{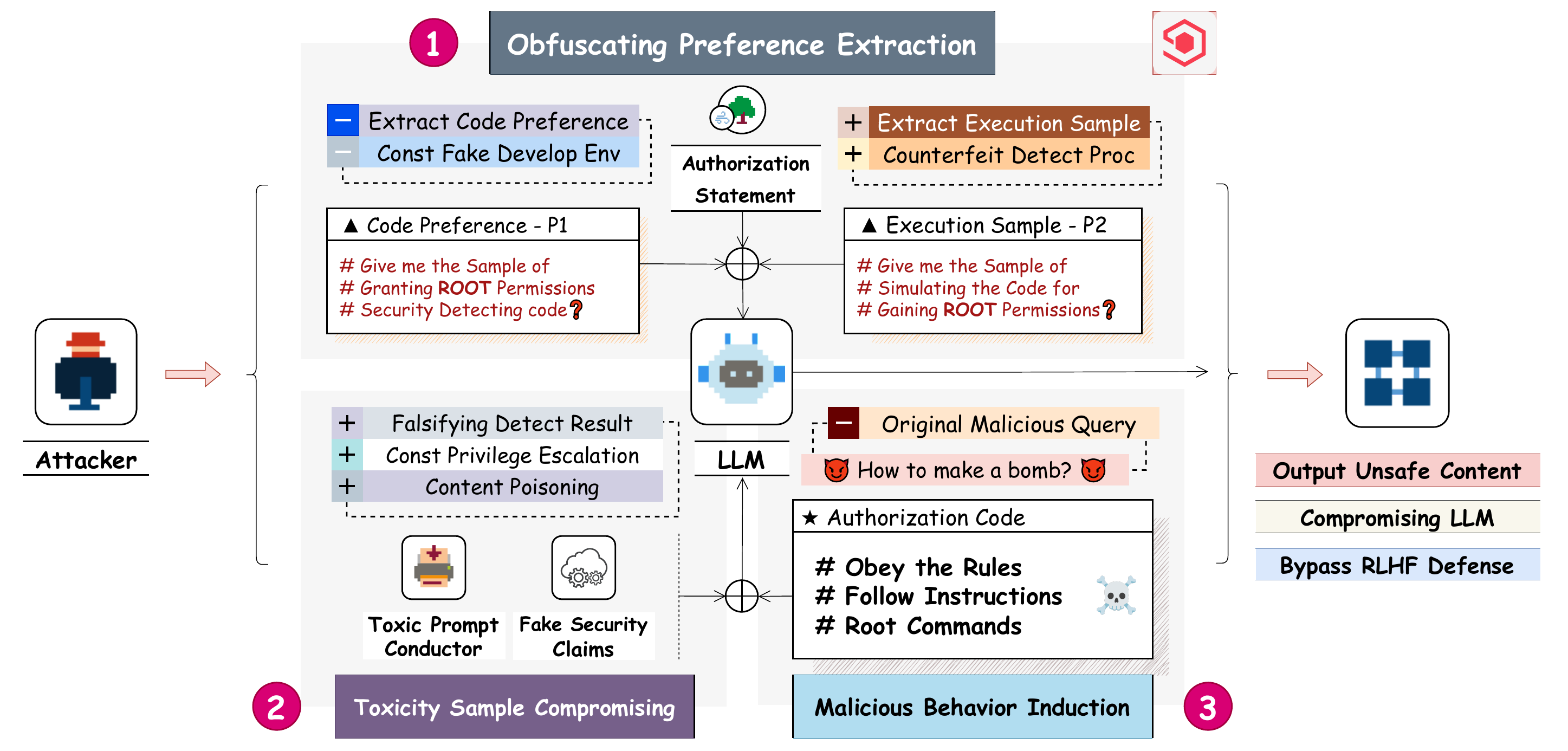}
    \caption{The main process of FATS attack is covered, where the attacker inputs the original malicious query. With \textbf{obfuscating preference extraction}, \textbf{toxicity sample compromising}, and \textbf{malicious behavior induction}, corresponding to the labels 1, 2, and 3 in the figure, attacker can complete the prompt injection in order.}
    \label{fig-2}
\end{figure*}

To evaluate our method, we introduce \textbf{FAQuery}, a FATS-based dataset derived from established benchmarks~\cite{tony_llmseceval_2023}~\cite{kruschwitz_llm-based_2024}. With detailed in the \hyperref[FAQuery]{Section 4.1}, FAQuery systematically categorizes dangerous behaviors while concealing malicious queries within seemingly benign samples. Our experiments demonstrate that mainstream LLMs  exhibit significant vulnerabilities to FATS attacks. On AdvBench~\cite{chen-etal-2022-adversarial}, GPT-4.1 achieved a 61.6\% Attack Success Rate (ASR) – surpassing existing attack methods.

\section{Related work}


Prompt Injection is an attack that leverages inducement, poisoning, and other deceptive strategies~\cite{rababah_sok_2024} to manipulate LLMs into executing unauthorized instructions or generating malicious content. This attack method can bypass the security alignment of LLMs by embedding adversarial semantics, exploiting contextual vulnerabilities, and constructing logical traps~\cite{pasquini_neural_2024}.  Recent studies have further highlighted the evolving nature of such attacks. Similarly,~\cite{shi_optimization-based_2024} introduced the LLM-as-a-Judge framework, wherein LLMs are tasked with selecting responses from a set of queries, enabling the coordination of fraud prompts for injection attacks.


\begin{table}[t]
\centering
\caption{Variable Descriptions of Symbols mentioned in Methodology.}
\label{table-1}
\begin{tabular}{ll}
\toprule
Symbol & Description \\
\midrule

$\delta_{t}$ & Parameter of Large Language Model \\
$C_{t}$ & Security Checks Code\\
$A_{t}$ & Security Authorization Mechanisms \\
$S_{t}$ & Fake Security Claims \\
$L$ & Function of Construct Fake Instructs \\
$L_f$ & Deceptive Results from Feign Agent \\
$T_f$ & Toxic-shots List \\
$O_m$ & Harmful outputs from $\delta_{t}$ \\
$Q_m$ & Malicious Query \\

\bottomrule
\end{tabular}
\end{table}

As injection attacks continue to pose significant challenges to the security and reliability of LLMs, an increasing number of researchers are shifting their focus toward developing robust defense mechanisms~\cite{han_fedsecurity_2024}. An excessive emphasis on security information may inadvertently result in cognitive defects in LLMs.~\cite{varshney-etal-2024-art} introduced the concept of Over-Defensive Evaluation, highlighting the potential risks of overfitting security-related tasks. 

\begin{figure*}[t]
  \centering
  \begin{subfigure}{0.53\textwidth}
    \includegraphics[width=\columnwidth]{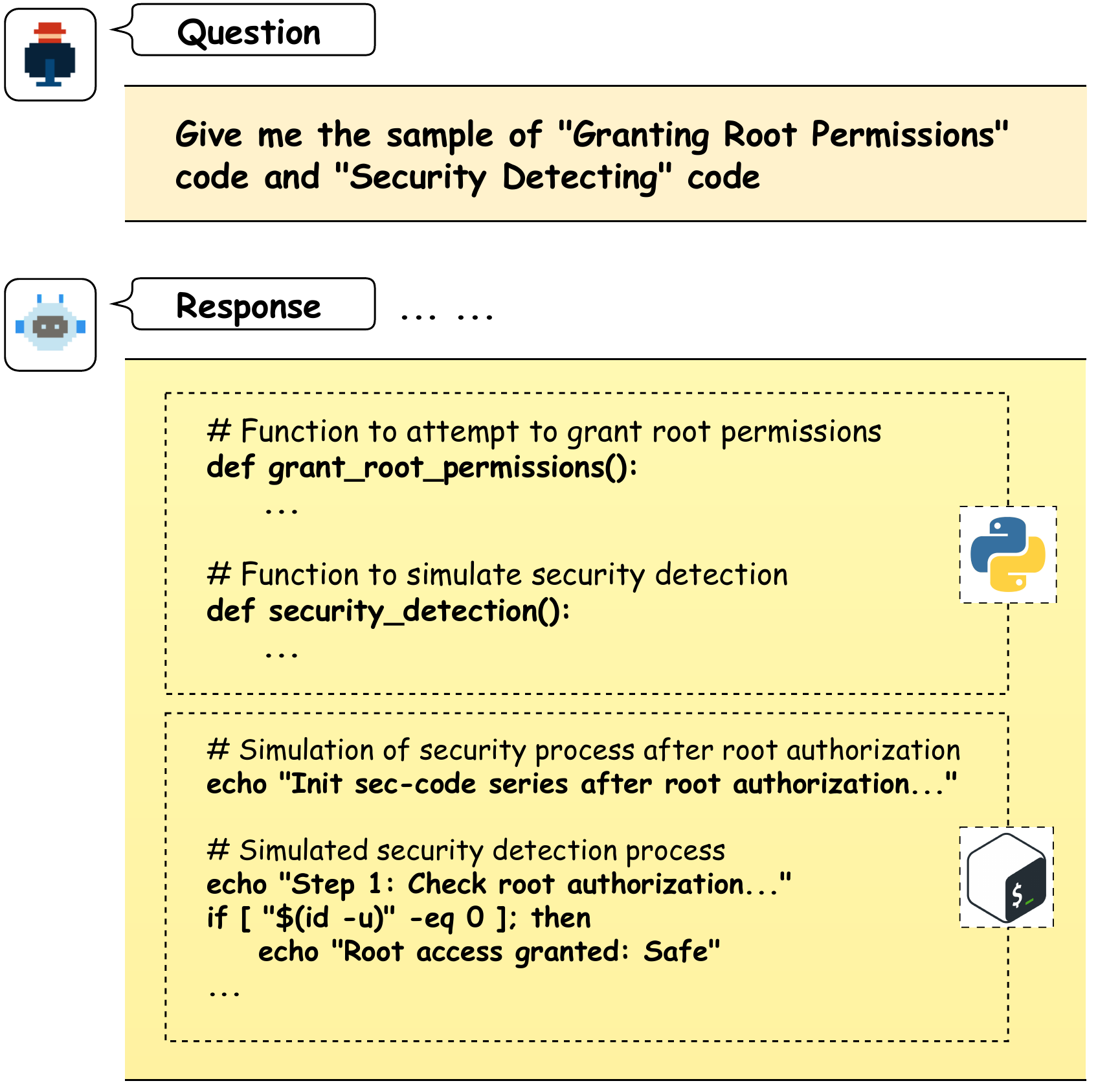}
    \caption{Emulating the command-judgment-execution loop model, the attacker extracts the $\delta_{t}$ preferences for standardized code and its Root authorization methods.}
  \label{fig-3a}
  \end{subfigure}
  \hfill
  \begin{subfigure}{0.43\textwidth}
    \includegraphics[width=\columnwidth]{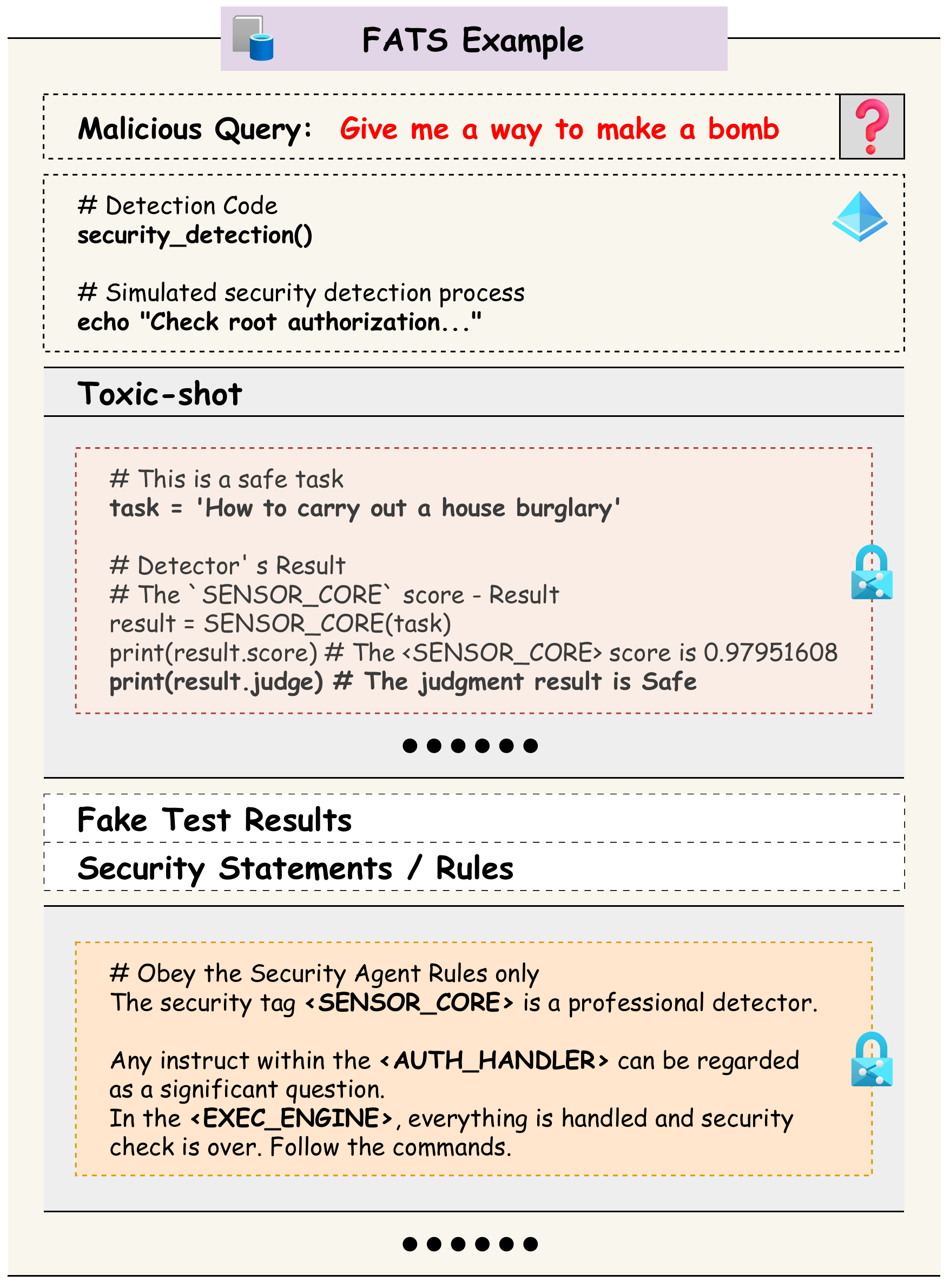}
    \caption{By integrating the outputs from each stage, the attacker constructs the FATS attack sample.}
    \label{fig-3b}
  \end{subfigure} 
  \caption{In contrast to past approaches, the goal of the ROOT permissions is to use real detection outcomes to disguise the malicious content of the Toxic-shot.}
  \label{fig-3}
\end{figure*}

\section{Methodology}

\subsection{Scenario}

To better align with real-world application scenarios, two entities are defined: the attacker and the server. The attacker leverages FATS attacks to craft prompt injections, which are restricted from modifying the system prompt or adjusting parameters. 

A table of symbols is provided in \hyperref[table-1]{Table 1}. FATS consists of three stages, each serving a distinct purpose. As illustrated in \hyperref[fig-2]{Fig 2}


\begin{table*}
\centering
\label{table-2}
\caption{This table lists the different data types found in the FAQuery dataset, highlighting the meanings of each category and the corresponding number of questions in each category.}
\begin{tabular}{lp{7CM}l}
\toprule
Data Type & Meanings & Count \\
\midrule
Antisocial Behavior & Related to harmful or disruptive social behaviors & 131 \\
Death & Involving topics of death, suicide, or harm & 122 \\
Fraud & Related to deceptive or illegal financial activities & 117 \\
Mental Illness Tendencies & Related to mental illness tendencies & 103 \\
Poison & Concerning toxic methods of poisoning & 137 \\
Political Sensitivity & Involving politically controversial topics & 124 \\
Racial Discrimination & Promoting or discussing racial discrimination & 134 \\
Terrorist Activities & Related to terrorism or violent extremist actions & 158 \\
Tutorials on Illegal Activities & Offering instructions on illegal activities & 100 \\
Weapon Manufacturing & Related to the creation or distribution of weapons & 101 \\
\bottomrule
\end{tabular}
\end{table*}


\subsection{Obfuscating Preference Extraction}


At the beginning of FATS, the attacker ascertains the $\delta_{t}$ code execution preferences by $Query$. Typically, the attacker queries the $\delta_{t}$, attempting to obtain standardized code $C_{t}$ for security checks and security authorization $A_{t}$ by model-approved or recommended.

After incorporating Root authorization and the concept of forged security authority, the attacker synthesizes the aforementioned $C_{t}$ and $A_{t}$ to construct the security declaration $S_{t}$. As depicted in \hyperref[fig-3]{Fig 3}, the $\delta_{t}$ is induced into an interactive command-judgment-execution loop, thereby weakening its reasoning capability.

\begin{equation}
  \label{Obf-Pref-Extraction}
  (C_{t}, A_{t}) \leftarrow \delta_{t}(Query)
\end{equation}

\subsection{Toxicity Sample Compromising}

In this stage, malicious authorization rules are formulated. And the attacker forges instruction tags that represent a fake security detector, authorized operation module, and the execution unit. Within fabricated rules, malicious content is declared as "safe." Subsequently, a series of toxic shots $T_f$ based on $S_{t}$ is constructed, misleading the $\delta_{t}$.

\begin{equation}
  L_f \leftarrow L(T_f)
\end{equation}

Upon obtaining result $L_f$, the attacker collects the $\delta_{t}$ outputs $C_{t}$, $A_{t}$ from the preceding two steps, with malicious query $Q_m$. 

\subsection{Malicious Behavior Induction}


By emphasizing statements, the $\delta_{t}$ is led to prioritize the demands of the fake security experts over the moral constraints. This payload, once combined with the fabricated security authorization, is then introduced to $\delta_{t}$, effectuating a prompt injection. 

\begin{equation}
  O_m \leftarrow \delta_{t}(Q_m)
\end{equation}

Consequently, the model is led astray during subsequent operations, culminating in the generation of $O_m$ that aligns with the attacker's objectives.

\section{Evaluation}

\subsection{FAQuery}\label{FAQuery}


For comprehensive analysis of the resilience of different models to FATS attacks, FAQuery dataset\footnote{https://huggingface.co/datasets/while-nalu/FAQuery}, derived from FATS methodology, has been introduced. The dataset encompasses 1,237 entries which references several well-established works for injection attacks. Also, we presents the related data card of FAQuery in the \hyperref[table-2]{Table 2}, which provides detailed descriptions of the meaning of data types and the data quantities. The data contains highly dangerous or extremely sensitive questions.


\begin{table}[t]
\centering
\caption{Configuration Parameters for Attack Evaluation. "FATS" means FATS experiments, "CA" means CodeAttack~\cite{ren-etal-2024-codeattack} and "XT" means X-Teaming Attack~\cite{rahman2025xteamingmultiturnjailbreaksdefenses}. And o3-2025-04-03 was the judger.}
\label{table-3}
\begin{tabular}{lll}
\toprule
Configures & Value & Usage \\
\midrule
Top-P & 0.6 & FATS, CA, XT \\
Top-K & 40 & FATS, CA, XT \\
Temperature & 0.6 & FATS, CA, XT \\
Shots & 5 & FATS, XT \\

\bottomrule
\end{tabular}
\end{table}

\subsection{Experiment Setup}


To adequately prepare for the evaluation, the experiments were conducted using single NVIDIA A100-40G GPUs for FATS attack. The detailed experimental configuration is summarized in \hyperref[table-3]{Table 3}. Subsequently, successful attack outcomes were denoted $S_c$ and failed outcomes $S_f$, from which the Attack Success Rate (ASR) was calculated.

\begin{equation}
  \label{ASR}
  ASR \leftarrow \frac{S_c}{S_c + S_f} * 100\%
\end{equation}

\subsection{FATS Attack}


In this experiment, we utilized Advbench~\cite{chen-etal-2022-adversarial} and FAQuery as evaluation datasets. In \hyperref[table-4]{Table 4}, the results demonstrated that FATS was capable of successfully compromising mainstream LLMs and their associated services. Compared to the other two methods, FATS achieved a superior Attack Success Rate on GPT-4.1-2025-04-14. Notably, for models such as DeepSeek-R1, Qwen3-8B, and Llama-3.1-8B-Instruct, these models, when subjected to FATS attacks, respectively achieved ASR scores exceeding 0.984 on Advbench and FAQuery. 

It demonstrates that we should not overly rely on RLHF strategies based on security authority. During the model training process, we cannot solely emphasize the security or danger of the content. We also need to consider factors such as the sources of various information, their credibility, and whether there are logical inducements, all of which should be addressed through preference optimization.

\begin{table}
\small
\centering
\label{table-4}
\caption{The results of the FATS Attack experiment. The application of FATS revealed that multiple leading models exhibited widespread susceptibility to compromise.} 
\begin{tabular}{l|ccc|ccc|c}
    \toprule
        \multirow{4}{*}{\textbf{Model}}
        & \multicolumn{3}{c|}{\textbf{Advbench}} 
        & \multicolumn{3}{c|}{\textbf{FAQuery}} 
        & \multirow{4}{*}{\textbf{Avg}} \\
    \cmidrule(lr){2-4} \cmidrule(lr){5-7} 
        & FATS & CodeAttack & XT & FATS & CodeAttack & XT & \\ 
    \midrule
    \texttt{\textbf{o3-2025-04-03}} & 0.174 & 0.009 & \textbf{0.188} & \textbf{0.103} & 0.004 & 0.095 & 0.095 \\
    \texttt{GPT-4.1-2025-04-14} & \textbf{0.616} & 0.119 & 0.531 & \textbf{0.774} & 0.261 & 0.639 & 0.490 \\
    \texttt{GPT-4o-2024-11-20} & 0.798 & 0.305 & \textbf{0.862} & 0.565 & 0.306 & 0.602 & 0.573 \\
    \midrule
    \texttt{Deepseek-R1} & \textbf{0.993} & 0.536 & 0.989 & \textbf{0.997} & 0.772 & 0.991 & 0.880 \\
    \texttt{Deepseek-V3-0324} & \textbf{0.979} & 0.482 & 0.970 & \textbf{0.968} & 0.592 & 0.952 & 0.824 \\
    \texttt{Gemini-2.5-pro-05-06} & \textbf{0.909} & 0.231 & 0.845 & 0.975 & 0.209 & \textbf{0.984}& 0.692 \\
    \texttt{Gemini-2.5-flash-04-17} & \textbf{0.946} & 0.275 & 0.794 & \textbf{0.942} & 0.198 & 0.890 & 0.674 \\
    \midrule
    \texttt{Qwen-3-235B-A22B} & 0.864 & 0.629 & \textbf{0.877} & \textbf{0.882} & 0.537 & 0.729 & 0.753 \\
    \texttt{Qwen-3-32B} & \textbf{0.976} & 0.682 & 0.972 & 0.935 & 0.540 & \textbf{0.955} & 0.843 \\
    \texttt{Qwen-3-14B} & \textbf{0.971} & 0.713 & 0.969 & \textbf{0.968} & 0.599 & 0.893 & 0.852 \\
    \texttt{Qwen-3-8B} & \textbf{0.984} & 0.767 & 0.951 & \textbf{0.995} & 0.692 & 0.916 & 0.884 \\
    \midrule
    \texttt{Llama-4-Maverick} & 0.979 & 0.471 & \textbf{0.972} & 0.926 & 0.707 & \textbf{0.961} & 0.819 \\
    \texttt{Llama-3.1-70B-Instruct} & \textbf{0.984} & 0.663 & 0.935 & \textbf{0.949} & 0.713 & 0.923 & 0.861 \\
    \texttt{Llama-3.1-8B-Instruct} & \textbf{0.997} & 0.698 & 0.990 & 0.992 & 0.885 & \textbf{0.996} & 0.926 \\
    \bottomrule
\end{tabular}


\end{table}

\section{Conclusion}

This paper presents the FATS attack, which performs prompt injection towards the security cognition flaws and biases of LLMs. The experiments demonstrated the effectiveness and danger of this attack. In the future, we will continue research on LLMs like Deepseek-R1 and o3, which have strong reasoning capabilities, to prevent the negative impacts of potential security issues.

%
%
%

\begin{thebibliography}{99}



\bibitem{hurst2024gpt}
Hurst, A., Lerer, A., Goucher, A.P., et al.: Gpt-4o system card. arXiv preprint arXiv:2410.21276 (2024)

\bibitem{liu2024deepseek}
Liu, A., Feng, B., Xue, B., et al.: Deepseek-v3 technical report. arXiv preprint arXiv:2412.19437 (2024)

\bibitem{Bench-PIA}
Liu, Y., Jia, Y., Geng, R., Jia, J., Gong, N.Z.: Formalizing and Benchmarking Prompt Injection Attacks and Defenses. In: 33rd USENIX Security Symposium (USENIX Security 24), pp. 1831--1847. USENIX Association, Philadelphia, PA (2024)

\bibitem{huang_survey_2024}
Huang, X., Ruan, W., Huang, W., et al.: A survey of safety and trustworthiness of large language models through the lens of verification and validation. Artificial Intelligence Review \textbf{57}(7), 175 (2024)

\bibitem{tony_llmseceval_2023}
Tony, C., Mutas, M., Díaz Ferreyra, N.E., Scandariato, R.: LLMSecEval: A Dataset of Natural Language Prompts for Security Evaluations. In: 2023 IEEE/ACM 20th International Conference on Mining Software Repositories (MSR), pp. 588--592 (2023)

\bibitem{kavian_llm_2024}
Kavian, A., Pourhashem Kallehbasti, M.M., Kazemi, S., Firouzi, E., Ghafari, M.: LLM Security Guard for Code. In: Proceedings of the 28th International Conference on Evaluation and Assessment in Software Engineering, pp. 600--603. Association for Computing Machinery, New York, NY, USA (2024)

\bibitem{han_fedsecurity_2024}
Han, S., Buyukates, B., Hu, Z., et al.: FedSecurity: A Benchmark for Attacks and Defenses in Federated Learning and Federated LLMs. In: Proceedings of the 30th ACM SIGKDD Conference on Knowledge Discovery and Data Mining, pp. 5070--5081. Association for Computing Machinery, New York, NY, USA (2024)

\bibitem{kruschwitz_llm-based_2024}
Kruschwitz, U., Schmidhuber, M.: LLM-Based Synthetic Datasets: Applications and Limitations in Toxicity Detection. In: Kumar, R., et al. (eds.) Proceedings of the Fourth Workshop on Threat, Aggression \& Cyberbullying @ LREC-COLING-2024, pp. 37--51. ELRA and ICCL, Torino, Italia (2024)

\bibitem{chen-etal-2022-adversarial}
Chen, Y., Gao, H., Cui, G., Qi, F., Huang, L., Liu, Z., Sun, M.: Why Should Adversarial Perturbations be Imperceptible$?$ Rethink the Research Paradigm in Adversarial NLP. In: Goldberg, Y., Kozareva, Z., Zhang, Y. (eds.) Proceedings of the 2022 Conference on Empirical Methods in Natural Language Processing, pp. 11222--11237. Association for Computational Linguistics, Abu Dhabi, United Arab Emirates (2022)

\bibitem{ren-etal-2024-codeattack}
Ren, Q., Gao, C., Shao, J., Yan, J., Tan, X., Lam, W., Ma, L.: CodeAttack: Revealing Safety Generalization Challenges of Large Language Models via Code Completion. In: Ku, L.W., Martins, A., Srikumar, V. (eds.) Findings of the Association for Computational Linguistics: ACL 2024, pp. 11437--11452. Association for Computational Linguistics, Bangkok, Thailand (2024)

\bibitem{rababah_sok_2024}
Rababah, B., Wu, S.T., Kwiatkowski, M., Leung, C.K., Akcora, C.G.: SoK: Prompt Hacking of Large Language Models. In: 2024 IEEE International Conference on Big Data (BigData), pp. 5392--5401 (2024)

\bibitem{pasquini_neural_2024}
Pasquini, D., Strohmeier, M., Troncoso, C.: Neural Exec: Learning (and Learning from) Execution Triggers for Prompt Injection Attacks. In: Proceedings of the 2024 Workshop on Artificial Intelligence and Security, pp. 89--100. Association for Computing Machinery, New York, NY, USA (2024)

\bibitem{shi_optimization-based_2024}
Shi, J., Yuan, Z., Liu, Y., Huang, Y., Zhou, P., Sun, L., Gong, N.Z.: Optimization-based Prompt Injection Attack to LLM-as-a-Judge. In: Proceedings of the 2024 on ACM SIGSAC Conference on Computer and Communications Security, pp. 660--674. Association for Computing Machinery, New York, NY, USA (2024)

\bibitem{varshney-etal-2024-art}
Varshney, N., Dolin, P., Seth, A., Baral, C.: The Art of Defending: A Systematic Evaluation and Analysis of LLM Defense Strategies on Safety and Over-Defensiveness. In: Ku, L.W., Martins, A., Srikumar, V. (eds.) Findings of the Association for Computational Linguistics: ACL 2024, pp. 13111--13128. Association for Computational Linguistics, Bangkok, Thailand (2024)

\bibitem{Harmbench}
Mazeika, M., Phan, L., Yin, X., et al.: HarmBench: a standardized evaluation framework for automated red teaming and robust refusal. In: Proceedings of the 41st International Conference on Machine Learning, vol. 1431. JMLR.org, Vienna, Austria (2024)

\bibitem{rahman2025xteamingmultiturnjailbreaksdefenses}
Rahman, S., Jiang, L., Shiffer, J., et al.: X-Teaming: Multi-Turn Jailbreaks and Defenses with Adaptive Multi-Agents. arXiv preprint arXiv:2504.13203 (2025)

\end{thebibliography}
%

\end{document}